\documentclass[10pt,english]{elsarticle}
\usepackage[T1]{fontenc}
\usepackage[latin9]{inputenc}
\usepackage{float}
\usepackage{amsmath}
\usepackage{amssymb}
\usepackage{stackrel}
\usepackage{graphicx}
\usepackage{wasysym}

\makeatletter
\usepackage{xcolor}
\newcommand{\pentg}{\scalebox{0.7}{\!\!\! $\rangle\!\!\!\sim\!\!\!\langle$}}
\newcommand{\spentg}{\scalebox{0.7}{ \!\!\!$\rangle\!\!\!\bumpeq\!\!\!\langle$}}
\newcommand{\tpentg}{\scalebox{0.7}{ \!\!\!$\rangle\!\!\!\Bumpeq\!\!\!\langle$}}
\newcommand{\upentg}{\scalebox{0.7}{ \!\!\!$\rangle\!\!\!\div\!\!\!\langle$}}
\newcommand{\dpentg}{\scalebox{0.7}{ \!\!\!$\rangle\!\!\!-\!\!\!\langle$}}
\newcommand{\vpentg}{\rotatebox[origin=c]{-90}{\pentg}}
\newcommand{\vspentg}{\rotatebox[origin=c]{-90}{\spentg}}
\newcommand{\vtpentg}{\rotatebox[origin=c]{-90}{\tpentg}}
\newcommand{\vupentg}{\rotatebox[origin=c]{-90}{\upentg}}
\newcommand{\vdpentg}{\rotatebox[origin=c]{-90}{\dpentg}}
\newcommand{\ptg}[4]{{}_{#1}^{#2}\pentg_{#4}^{#3}}

\newcommand{\vptg}[4]{{}_{#1}^{#2}\vpentg_{#4}^{#3}}

\usepackage{babel}

\makeatother

\usepackage{babel}
\begin{document}
\begin{frontmatter}{}

\title{Geometrically frustrated Cairo pentagonal lattice stripe with Ising
and Heisenberg exchange interactions}

\author[{*}]{F. C. Rodrigues, S. M. de Souza and Onofre Rojas}

\cortext[a]{email: ors@dfi.ufla.br}

\address{Departamento de Física, Universidade Federal de Lavras, CP 3037,
37200-000, Lavras-MG, Brazil}
\begin{abstract}
Motivated by the recent discoveries of some compounds such as the
$\mathrm{Bi}_{2}\mathrm{Fe}_{4}\mathrm{O}_{9}$ which crystallizes
in an orthorhombic crystal structure with the $\mathrm{Fe}^{3+}$
ions, and iron-based oxyfluoride ${\rm Bi}_{4}{\rm Fe}_{5}{\rm O}_{13}{\rm F}$
compounds following the pattern of Cairo pentagonal structure, among
some other compounds. We propose a model for one stripe of the Cairo
pentagonal Ising-Heisenberg lattice, one of the edges of a pentagon
is different, and this edge will be associated with a Heisenberg exchange
interaction, while the Ising exchange interactions will associate
the other edges. We study the phase transition at zero temperature,
illustrating five phases: a ferromagnetic phase (FM), a dimer antiferromagnetic
(DAF), a plaquette antiferromagnetic (PAF), a typical antiferromagnetic
(AFM) and a peculiar frustrated phase (FRU) where two types of frustrated
states with the same energy coexist. To obtain the partition function
of this model, we use the transfer matrix approach and following the
eight vertex model notation. Using this result we discuss the specific
heat, internal energy and entropy as a function of the temperature,
and we can observe some unexpected behavior in the low-temperature
limit, such as anomalous double peak in specific heat due to the existence
of three phase (FRU, PAF(AFM) and FM) transitions occurring in a close
region to each other. Consequently, the low-lying energy thermal excitation
generates this double anomalous peak, and we also discuss the internal
energy at the low temperature limit, where this double peak curve
occurs. Some properties of our result were compared with two dimensional
Cairo pentagonal lattices, as well as orthogonal dimer plaquette Ising-Heisenberg
chain. 
\end{abstract}
\end{frontmatter}{}

\section{Introduction}

In the past few years, the investigation in the Cairo pentagonal lattice
has drawn much attention to researchers in condensed matter physics.
The first model proposed with the Cairo pentagonal Ising lattice structure
was published in 2002 by Urumov\citep{urumov}, where the author studied
as a purely theoretical physics problem, years later motivating a
significant impact in pentagonal lattice investigation. More recently,
the geometric frustration of Cairo pentagonal Ising model also was
studied in more detail in reference \citep{Mrojas}. By the year of
2009 Ressouche et al.\citep{ressou} identified a compound $\mathrm{Bi}_{2}\mathrm{Fe}_{4}\mathrm{O}_{9}$
which crystallizes in an orthorhombic structure with the $\mathrm{Fe}^{3+}$
ions, forming a first analogue with a magnetic Cairo pentagonal tessellation,
where the Heisenberg exchange interaction describes the couplings
with a good approximation. Furthermore, Ralko \citep{ralko} studied
the phase diagram of Cairo Pentagonal XXZ with spin-1/2 under a magnetic
field, discussing the zero and finite temperature properties, using
the stochastic series expansion and the cluster mean-field theory
approach. Next, Rousochatzakis et al.\citep{Rouso} performed an extensive
investigation both analytically and numerically, for the antiferromagnetic
Heisenberg model on the Cairo pentagonal lattice. Following, Pchelkina
and Streltsov\citep{pchelkina} investigated the electronic structure
and magnetic properties of compound $\mathrm{Bi}_{2}\mathrm{Fe}_{4}\mathrm{O}_{9}$,
forming a Cairo pentagonal lattice with strong geometric frustration.
Besides, Abakumov et al.\citep{abakumov} reported a new crystal structure
and magnetism of the iron-based oxyfluoride ${\rm Bi}_{4}{\rm Fe}_{5}{\rm O}_{13}{\rm F}$,
and this compound also exhibits a Cairo pentagonal structure. Later,
Isoda et al.\citep{Isoda} also studied the magnetic phase diagram
under a magnetic field\citep{Isoda} as well as its magnetization
process\citep{nakano} of the spin-1/2 Heisenberg antiferromagnetic
on the Cairo pentagonal lattice. More recently, a novel compound ${\rm Bi}_{2}{\rm Fe}_{4-x}{\rm Cr}_{x}{\rm O}_{9}$
($x=0.5,1,1.2$) has been synthesized using a soft chemistry technique
followed by a solid-state reaction in Ar\citep{Rozova}, which is
a highly homogeneous mullite-type solid.

There are even more new compounds studied with the same structure,
such the 2D crystals $\mathrm{SnX}_{2}$ (X = S, Se, or Te)\citep{Ma}
which have been investigated using a first-principle calculation.
Another investigation was carried out by Chainani and Sheshadri \citep{chainani}
for the Ising model with Cairo pentagonal pattern with a nearest-neighbor
antiferromagnetic coupling. Finally, we can still comment on the new
penta-graphene, recently discovered by Zhang et al.\citep{Zhang},
although this compound is a non magnetic one, the penta-graphene pattern
follows exactly the same Cairo pentagonal tessellation.

On the other hand, the Cairo pentagonal lattice Ising model\citep{urumov,Mrojas}
is the dual of Shastry-Sutherland lattice Ising model\citep{shastry}.
Motivated by the Shastry-Sutherland lattice, Ivanov proposed a quasi
one-dimensional Heisenberg model called as orthogonal dimer plaquette
chain\citep{ivanov}. Certainly, the Cairo pentagonal chain can be
viewed as a decorated orthogonal dimer chain\citep{henrique,taras},
where in our case the Ising spin would be considered as decorated
spins. However, we cannot use the decoration transformation approach\citep{dec-trans}
naively to map the Cairo pentagonal chain into an orthogonal dimer
chain, because we have quantum spins instead of classical spins (required
condition to apply decoration transformation technique). Although
there is a proposal for quantum spin decoration transformation approach\citep{braz},
this transformation is exact only for isolated decorations, applying
to a quantum spin lattice model would be just an approximate mapping.

The outline of this work is as follows. In sec. 2, we present the
pentagonal Ising-Heisenberg chain, and we discuss the phase diagram
at zero temperature. In sec. 3, we present the details to obtain the
free energy calculation, and in sec. 4, we discuss some physical quantities
obtained from free energy, such as the entropy, specific heat and
internal energy. Finally, in sec. 5 we summarize our results and draw
our conclusions.

\section{Cairo pentagonal Ising-Heisenberg lattice stripe}

\begin{figure}[h]
\centering{}\includegraphics[scale=0.71]{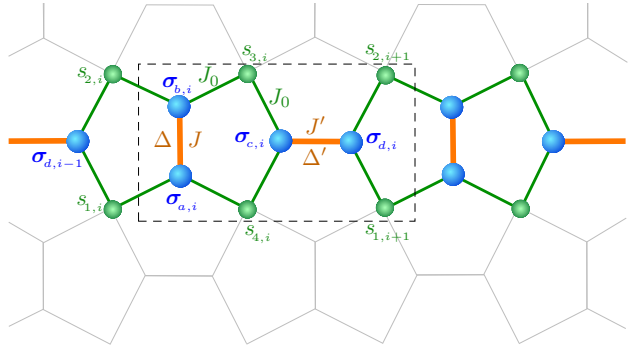}\caption{\label{fig:pent-chain}Schematic representation of the Cairo pentagonal
Ising-Heisenberg lattice stripe, $\sigma$ represents Heisenberg spins
and $s$ represents the Ising spins. The dashed rectangle represents
an unit cell. }
\end{figure}

Motivated by the comments given in the introduction we consider a
stripe of the Cairo pentagonal lattice or decorated orthogonal dimer
plaquette chain with Ising-Heisenberg coupling as schematically depicted
in fig. \ref{fig:pent-chain}. 

Let us define that the Hamiltonian for a Cairo pentagonal chain by
\begin{equation}
H=\sum_{i=1}^{N}\left(H_{i}^{ab}+H_{i,i+1}^{cd}\right),\label{eq:Hamilt-orig}
\end{equation}
where $N$ is the number of cells and assuming periodic boundary condition.
Thus, let us call as \textquotedbl{}elementary cell\textquotedbl{}
$ab$-dimer and $cd$-dimer, which are described by $H_{i}^{ab}$
and $H_{i,i+1}^{cd}$ respectively.

Therefore, each block Hamiltonian become
\begin{alignat}{1}
H_{i}^{ab}= & -J(\boldsymbol{\sigma}_{a,i},\boldsymbol{\sigma}_{b,i})_{\Delta}-J_{0}(s_{1,i}+s_{4,i})\sigma_{a,i}^{z}+\nonumber \\
 & -J_{0}(s_{2,i}+s_{3,i})\sigma_{b,i}^{z},\label{eq:Ham-ab}\\
H_{i,i+1}^{cd}= & -J'(\boldsymbol{\sigma}_{c,i},\boldsymbol{\sigma}_{d,i})_{\Delta'}-J_{0}(s_{3,i}+s_{4,i})\sigma_{c,i}^{z}+\nonumber \\
 & -J_{0}(s_{1,i+1}+s_{2,i+1})\sigma_{d,i}^{z},\label{eq:Ham-cd}
\end{alignat}
here $\sigma_{\gamma,i}^{\alpha}$ are the spin operators (with $\alpha=\{x,y,z\}$)
at site $i$ for particles $\gamma=\{a,b,c,d\}$, for detail see fig.\ref{fig:pent-chain}.
The Ising spin exchange interaction parameter is denoted by $J_{0}$,
whereas $J$ ($J'$) represents the Heisenberg exchange interaction
and $\Delta$ ($\Delta'$) means the anisotropic exchange interaction
between Heisenberg spins for $ab$-dimer ($cd$-dimer) respectively.
Whereas for $ab$-dimer $J(\boldsymbol{\sigma}_{a,i},\boldsymbol{\sigma}_{b,i})_{\Delta}$
is defined by
\begin{alignat}{1}
J(\boldsymbol{\sigma}_{a,i},\boldsymbol{\sigma}_{b,i})_{\Delta}\equiv & J(\sigma_{a,i}^{x}\sigma_{b,i}^{x}+\sigma_{a,i}^{y}\sigma_{b,i}^{y})+\Delta\sigma_{a,i}^{z}\sigma_{b,i}^{z},\label{eq:J-Dlt}
\end{alignat}
and for $cd$-dimer $J'(\boldsymbol{\sigma}_{a,i},\boldsymbol{\sigma}_{b,i})_{\Delta'}$
is defined analogously to eq.\eqref{eq:J-Dlt}.

\subsection{Zero temperature phase diagram}

To study the phase diagram at zero temperature, we need to describe
the ground-state energy per unit cell. One elementary cell is composed
by one dimer and bonded to 4 Ising spins. Each unit cell can be composed
by two elementary cells: one $ab$-dimer and one $cd$-dimer, both
dimers are bonded by 2 Ising spins. In fig.\ref{fig:pent-chain} is
illustrated one possible unit cell represented by a dashed rectangle.

The ground-state energy for each elementary cell could be described
schematically using the fancy notations by

\begin{alignat}{1}
ab-\text{dimer }\longrightarrow\: & |\vptg{s_{1}}{s_{2}}{s_{3}}{s_{4}}\rangle\;\text{and}\\
cd-\text{dimer }\longrightarrow\: & |\ptg{s_{4}}{s_{3}}{s_{5}}{s_{6}}\rangle,
\end{alignat}
where $s_{i}$ correspond to the Ising spins, while the fancy symbols
$\vptg{}{}{}{}$ and $\ptg{}{}{}{}$ denote four possible states as
a function of $\{s_{i}\}$. Here, we use the spin subindex just for
convenience, which cannot be confused with a more explicit spins notation
in the Hamiltonian \eqref{eq:Ham-ab} and \eqref{eq:Ham-cd}. Therefore,
the eigenstates $|\vptg{s_{1}}{s_{2}}{s_{3}}{s_{4}}\rangle$ of $ab$-dimers
are conveniently expressed using four additional fancy notations,
which are represented as follows

\begin{alignat}{1}
|\vupentg\rangle= & |_{+}^{+}\rangle,\label{eq:uu}\\
|\vspentg\rangle= & \bigl(-\sin(\tfrac{\phi}{2})|_{-}^{+}\rangle+\cos(\tfrac{\phi}{2})|_{+}^{-}\rangle\bigr),\label{eq:sngl}\\
|\vtpentg\rangle= & \bigl(\cos(\tfrac{\phi}{2})|_{-}^{+}\rangle+\sin(\tfrac{\phi}{2})|_{+}^{-}\rangle\bigr),\label{eq:trip}\\
|\vdpentg\rangle= & |_{-}^{-}\rangle,\label{eq:dd}
\end{alignat}
where $\phi=\arctan\bigl(\tfrac{J}{J_{0}(s_{1}-s_{2}-s_{3}+s_{4})}\bigr)$,
with $-\pi\leqslant\phi\leqslant\pi$.

The first eigenstate $|_{+}^{+}\rangle$ of the $ab$-dimer \eqref{eq:uu}
is denoted by $|\vupentg\rangle$, which is linked to spins $\{s_{1},s_{2},s_{3},s_{4}\}$,
thats why we use this fancy notation. Similarly, the eigenstate \eqref{eq:dd}
is denoted by $|\vdpentg\rangle$ which represents the states $|_{-}^{-}\rangle$
linked to the same set of spins $\{s_{1},s_{2},s_{3},s_{4}\}$. Whereas,
$|\vspentg\rangle$ denotes eq.\eqref{eq:sngl} some kind of \textquotedbl{}anti-symmetric\textquotedbl{}
state for $\phi>0$, and $|\vtpentg\rangle$ represents \eqref{eq:trip}
some kind of \textquotedbl{}symmetric\textquotedbl{} state for $\phi>0$,
although for $\phi<0$ this affirmation exchanges.

It is worth to mention that the states $|\vupentg\rangle$ and $|\vdpentg\rangle$
are independent of Ising spins $\{s_{1},s_{2},s_{3},s_{4}\}$, whereas
states $|\vspentg\rangle$ and $|\vtpentg\rangle$ depends of $\phi$
which subsequently depends of Ising spins $\{s_{1},s_{2},s_{3},s_{4}\}$.

Whereas the corresponding energy eigenvalues for $ab$-dimer are given
by
\begin{alignat}{2}
\vupentg\;\mapsto\quad & \epsilon_{1}= &  & -\tfrac{\Delta}{4}+\tfrac{J_{0}}{2}(s_{1}+s_{4}+s_{2}+s_{3}),\label{eq:Ee1}\\
\vspentg\;\mapsto\quad & \epsilon_{2}= &  & \tfrac{\Delta}{4}+\tfrac{1}{2}\sqrt{J_{0}^{2}(s_{1}+s_{4}-s_{2}-s_{3})^{2}+J^{2}},\\
\vtpentg\;\mapsto\quad & \epsilon_{3}= &  & \tfrac{\Delta}{4}-\tfrac{1}{2}\sqrt{J_{0}^{2}(s_{1}+s_{4}-s_{2}-s_{3})^{2}+J^{2}},\\
\vdpentg\;\mapsto\quad & \epsilon_{4}= &  & -\tfrac{\Delta}{4}-\tfrac{J_{0}}{2}(s_{1}+s_{4}+s_{2}+s_{3}).\label{eq:Ee4}
\end{alignat}

Now using the eight-vertex model notation\citep{urumov,Mrojas,baxter}
and the fancy notations (\ref{eq:uu}-\ref{eq:dd}), we define six
different states, explicitly including the Ising spins that connect
the elementary cells by
\begin{alignat}{2}
|u_{1}\rangle=|\vptg{+}{+}{+}{+}\rangle,\quad & |u_{2}\rangle=|\vptg{+}{-}{+}{-}\rangle,\quad & |u_{3}\rangle=|\vptg{+}{-}{-}{+}\rangle,\nonumber \\
|u_{4}\rangle=|\vptg{+}{+}{-}{-}\rangle,\quad & |u_{5}\rangle=|\vptg{+}{-}{+}{+}\rangle,\quad & |u_{7}\rangle=|\vptg{+}{+}{+}{-}\rangle.\label{eq:u_i}
\end{alignat}
The reason why we choose this notation could be more evident in the
next section.

To obtain all 16 possible states, we can use in relation \eqref{eq:u_i}
the vertical and spin inversion symmetry, to recover 10 remaining
states. Whereas each corresponding elementary cell ($|u_{i}\rangle$)
energy is defined by $\epsilon_{i}$, which have the following properties
$\epsilon_{2}=\epsilon_{4}$ and $\epsilon_{5}=\epsilon_{6}=\epsilon_{7}=\epsilon_{8}$.
Note that each $|u_{i}\rangle$ represents symbolically four states
given by (\ref{eq:uu}-\ref{eq:dd}).

Analogously, we can obtain the corresponding states $|\ptg{s_{4}}{s_{3}}{s_{5}}{s_{6}}\rangle$,
by the same relation to that (\ref{eq:Ee1}-\ref{eq:Ee4}), consequently
the eigenvalues are expressed merely by substituting $J\rightarrow J'$,
$\Delta\rightarrow\Delta'$ and $\{s_{1},s_{2},s_{3},s_{4}\}\rightarrow\{s_{6},s_{4},s_{3},s_{5}\}$. 

Thus, the energy eigenvalues become 
\begin{alignat}{2}
\upentg\;\mapsto\; & \epsilon'_{1}= &  & -\tfrac{\Delta'}{4}+\tfrac{J_{0}}{2}(s_{6}+s_{5}+s_{4}+s_{3}),\label{eq:Ee1-1}\\
\spentg\;\mapsto\; & \epsilon'_{2}= &  & \tfrac{\Delta'}{4}+\tfrac{1}{2}\sqrt{J_{0}^{2}(s_{6}+s_{5}-s_{4}-s_{3})^{2}+{J'}^{2}},\\
\tpentg\;\mapsto\; & \epsilon'_{3}= &  & \tfrac{\Delta'}{4}-\tfrac{1}{2}\sqrt{J_{0}^{2}(s_{6}+s_{5}-s_{4}-s_{3})^{2}+{J'}^{2}},\\
\dpentg\;\mapsto\; & \epsilon'_{4}= &  & -\tfrac{\Delta'}{4}-\tfrac{J_{0}}{2}(s_{6}+s_{5}+s_{4}+s_{3}).\label{eq:Ee4-1}
\end{alignat}
Substituting $\phi\rightarrow\phi'$, the corresponding eigenstates
read as follows 
\begin{alignat}{1}
|\,\upentg\rangle= & |++\rangle,\\
|\,\spentg\rangle= & -\sin(\tfrac{\phi'}{2})|+-\rangle+\cos(\tfrac{\phi'}{2})|-+\rangle,\\
|\,\tpentg\rangle= & \cos(\tfrac{\phi'}{2})|+-\rangle+\sin(\tfrac{\phi'}{2})|-+\rangle,\\
|\,\dpentg\rangle= & |--\rangle,
\end{alignat}
where $\phi'=\arctan\bigl(\tfrac{J'}{J_{0}(s_{6}+s_{5}-s_{4}-s_{3})}\bigr)$,
with $-\pi\leqslant\phi'\leqslant\pi$. 

Similarly, the corresponding elementary states $|u_{j}\rangle\rightarrow|\bar{u}_{j}\rangle$
can be obtained by rotating in $\pi/2$ all 16 states. Then the first
rotated state becomes $|\bar{u}_{1}\rangle=|\ptg{+}{+}{+}{+}\rangle$,
thus all other states could be similarly obtained.

After defining each elementary cell, now we can construct the unit
cell states by $|u_{j}\rangle\otimes|\bar{u}_{k}\rangle$. Once more,
let us use the eight-vertex model notation\citep{urumov,Mrojas} to
simplify the eigenstate of the unit cell, 
\begin{equation}
|v_{i}\rangle=|\vptg{s_{1}}{s_{2}}{s_{3}}{s_{4}}\ptg{}{}{s'_{2}}{s'_{1}}\rangle.\label{eq:rot-u}
\end{equation}
Here becomes useful the fancy notation, because it relates the unit
cell structure clearly. The unit cell state denoted by $|v_{i}\rangle$
are closely related with eight-vertex model notation for Ising spins
$\{s_{1},s_{2},s'_{2},s'_{1}\}$, here again we denote the Ising spins
only for convenience by $s'_{1}=s_{6}$ and $s'_{2}=s_{5}$. Thus,
the eq.\eqref{eq:rot-u} can be expressed as follows
\begin{alignat}{2}
|v_{1}\rangle= & |\vptg{+}{+}{s_{3}}{s_{4}}\ptg{}{}{+}{+}\rangle,\quad & |v_{2}\rangle= & |\vptg{+}{-}{s_{3}}{s_{4}}\ptg{}{}{+}{-}\rangle,\nonumber \\
|v_{3}\rangle= & |\vptg{+}{-}{s_{3}}{s_{4}}\ptg{}{}{-}{+}\rangle, & |v_{4}\rangle= & |\vptg{+}{+}{s_{3}}{s_{4}}\ptg{}{}{-}{-}\rangle,\nonumber \\
|v_{5}\rangle= & |\vptg{+}{-}{s_{3}}{s_{4}}\ptg{}{}{+}{+}\rangle, & |v_{7}\rangle= & |\vptg{+}{+}{s_{3}}{s_{4}}\ptg{}{}{+}{-}\rangle.\label{eq:v-st}
\end{alignat}

It is worth mentioning that the unit cell convention allows expressing
the eigenstates, this means that the edge of a unit cell must always
connect the spins that share two neighboring unit cells. Thus, to
satisfy the number of particles per unit cell, the leftmost and rightmost
Ising spins must be shared by two unit cells, then each shared particle
contributes with a half \textquotedbl{}particle\textquotedbl{} in
the unit cell, as described by the fancy notations.

Certainly, each unit cell state $|v_{i}\rangle$ represents symbolically
the $4\times4\times4=64$ possible states, the most relevant states
are given by \eqref{eq:v-st} and the remaining configurations can
be obtained using horizontal symmetry and spin inversion symmetry.

Nevertheless, the rotational symmetry and the vertical symmetry are
not allowed, because $|\vptg{}{}{}{}\;\ptg{}{}{}{}\rangle\;\ne\;|\ptg{}{}{}{}\;\vptg{}{}{}{}\rangle$,
this means that the local chiral symmetry is broken in each unit cell,
although the global chiral symmetry is preserved.

Using the previous result, we can study the phase diagram of the ground-state
energy for the Cairo pentagonal chain per unit cell, thus we obtain
\begin{alignat}{1}
E_{_{FM}}= & -2J_{0}-\frac{\Delta}{2},\\
E_{_{AFM}}= & J_{0}-\tfrac{1}{2}\sqrt{J^{2}+4J_{0}^{2}},\\
E_{_{PAF}}= & -J_{0}-\tfrac{1}{2}\sqrt{J^{2}+4J_{0}^{2}},\\
E_{_{DAF}}= & 2J_{0}-\frac{\Delta}{2},\\
E_{_{FRU}}= & \tfrac{1}{2}\Delta-\tfrac{1}{2}|J|-\tfrac{1}{2}\sqrt{J^{2}+4J_{0}^{2}},\label{eq:E-fru}
\end{alignat}
where we consider for simplicity $J'=J$ and $\Delta'=\Delta$. 

Thus the system exhibits five states, whose ground-states can be expressed
by
\begin{alignat}{1}
|FM\rangle= & \prod_{i=1}^{N}|_{+}^{+}\,\vupentg\,{}_{+}^{+}\,\upentg\,{}_{+}^{+}\rangle_{i}\quad\text{or}\quad\prod_{i=1}^{N}|_{-}^{-}\,\vdpentg\,|_{-}^{-}\,\dpentg\,|_{-}^{-}\rangle_{i},\label{eq:fm}\\
|DAF\rangle= & \prod_{i=1}^{N}|_{+}^{+}\,\vdpentg\,{}_{+}^{+}\,\dpentg\,{}_{+}^{+}\rangle_{i},\quad\text{or}\quad\prod_{i=1}^{N}|_{-}^{-}\,\vupentg\,|_{-}^{-}\,\upentg\,|_{-}^{-}\rangle_{i},\\
|PAF\rangle= & \prod_{i=1}^{N/2}|_{+}^{+}\,\vupentg\,{}_{+}^{+}\,\tpentg\,{}_{-}^{-}\,\vdpentg\,{}_{-}^{-}\,\tpentg\,{}_{+}^{+}\rangle_{i},\\
|AFM\rangle= & \prod_{i=1}^{N/2}|_{+}^{+}\,\vdpentg\,{}_{+}^{+}\,\tpentg\,{}_{-}^{-}\,\vupentg\,{}_{-}^{-}\,\tpentg\,{}_{+}^{+}\rangle_{i}.\label{eq:afm}
\end{alignat}

The Cairo pentagonal Ising-Heisenberg chain can be described by the
Hamiltonian \eqref{eq:Hamilt-orig}, which exhibits five states (\ref{eq:fm}-\ref{eq:afm}):
where we found a ferromagnetic ($FM$) phase; three types of antiferromagnetic
phase, a dimer antiferromagnetic (DAF) phase, a plaquette antiferromagnetic
(PAF) phase and one antiferromagnetic (AFM); Surely, the states (\ref{eq:fm}-\ref{eq:afm})
satisfy the spin inversion symmetry, all Ising and Heisenberg spins
inversion leaves the system invariant. 

The other state corresponding to the energy \eqref{eq:E-fru} is frustrated
($FRU$), represented symbolically by 

\begin{alignat}{1}
|FRU\rangle= & \left\{ \begin{array}{lc}
\stackrel[i=1]{N}{\prod}\negmedspace|{}_{\tau_{i-1}}^{\tau_{i-1}}\,\vtpentg\,{}_{-\tau_{i}}^{-\tau_{i}}\,\tpentg\,{}_{\tau_{i}}^{\tau_{i}}\rangle_{i}, & \rightarrow\;\text{Frustration type I,}\\
\stackrel[i=1]{N}{\prod}\negmedspace|{}_{\tau_{i}}^{-\tau_{i}}\,\vtpentg\,{}_{\tau_{i}}^{-\tau_{i}}\,\tpentg\,{}_{\tau_{i+1}}^{-\tau_{i+1}}\rangle_{i}, & \rightarrow\;\text{Frustration type II},
\end{array}\right.\label{eq:fru}
\end{alignat}
where $\tau_{i}$ ($\tau_{i,i+1}$) can take independently $\pm$
in each unit cell. We can recognize a frustrated state of type I is
degenerate in $2^{N}$ states, since for each unit cell there are
2 degrees freedom \eqref{eq:fru}. Similarly, for the frustration
of type II becomes $2^{N}$ possible configurations (states) \eqref{eq:fru}.
Therefore, in total we have $2\times2^{N}$ degenerate states. Notice
that frustration type I and II cannot be mixed, because the linking
spins of each unit cells are incompatible. Thus, we find a residual
entropy $\mathcal{S}=k_{B}\ln(2\times2^{N})/N=k_{B}\ln(2)$. It is
worth remembering that the factor $2$ that multiplies $2^{N}$ corresponds
to two type of frustrations, but in thermodynamic limit this factor
becomes irrelevant, this peculiar property is unusual for frustrated
systems. 

\begin{figure}
\centering{}\includegraphics[scale=0.22]{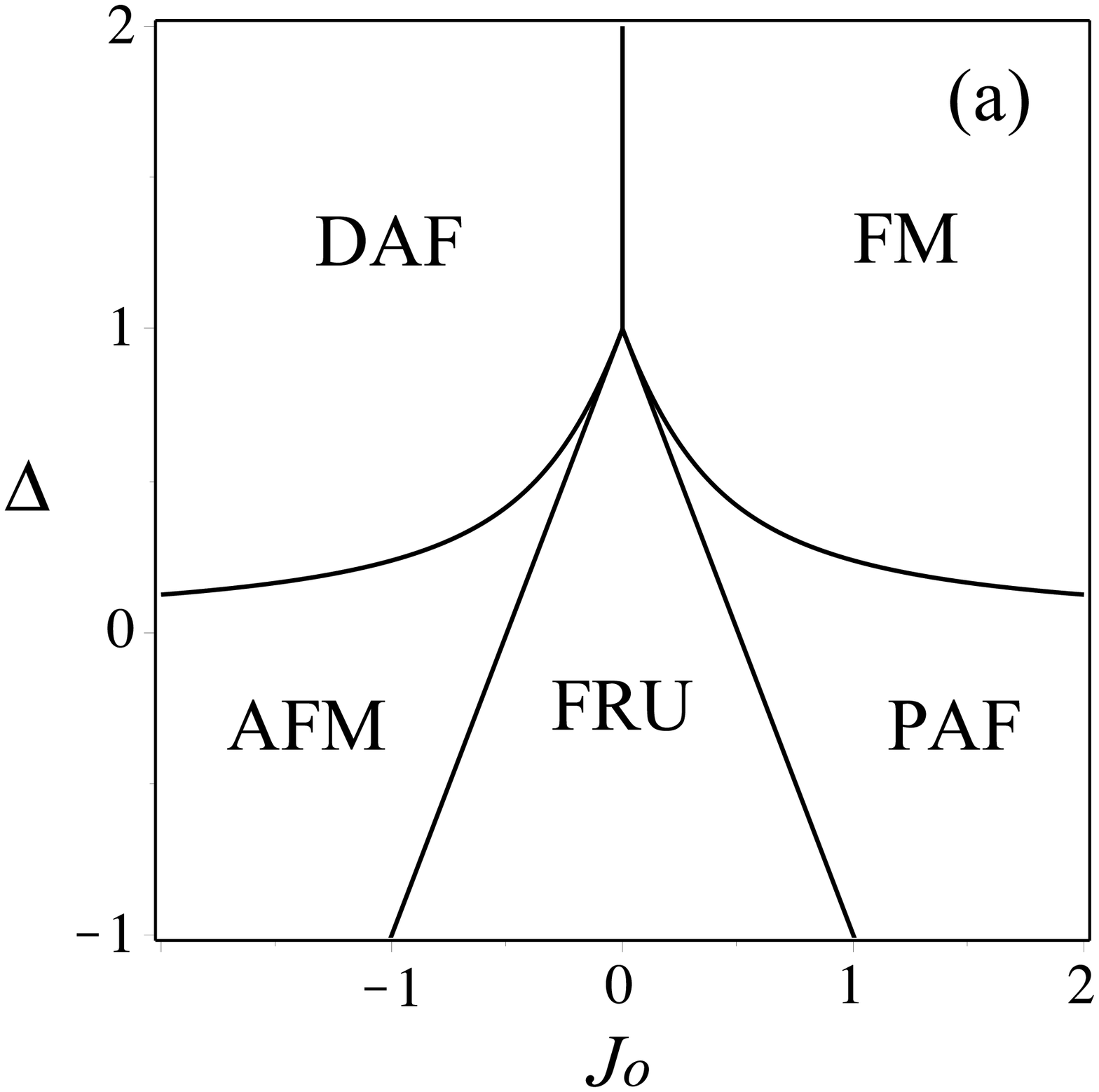}\includegraphics[scale=0.22]{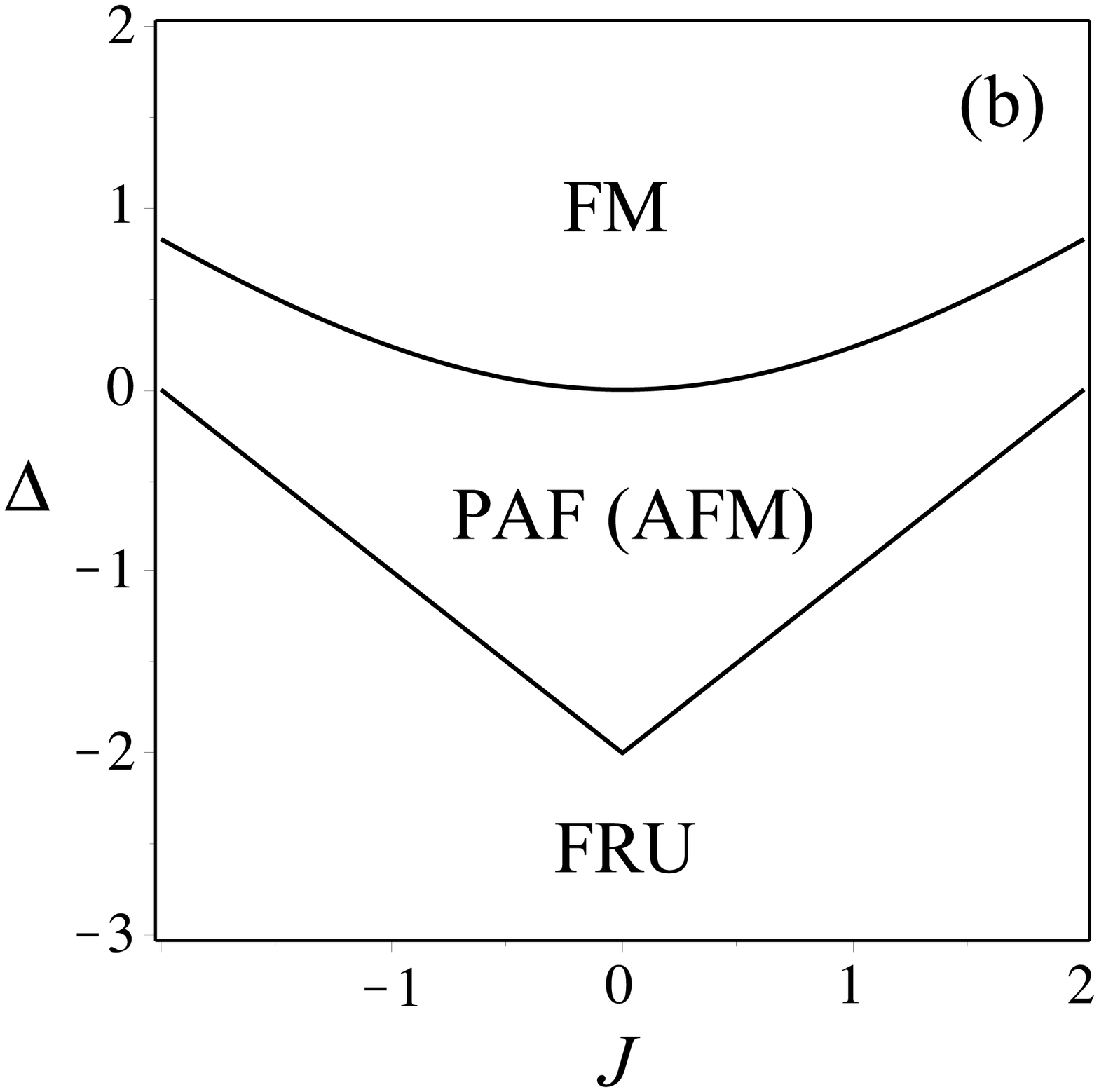}\caption{\label{fig:Phase-diagram-zero}Ground-state phase diagram, where is
illustrated a ferromagnetic phase ($FM$), a plaquette antiferromagnetic
($PAF$), a dimer antiferromagnetic ($DAF$) and a frustrated (FRU)
phase. (a) In plane $J_{0}-\Delta$, for fixed $J=1$. (b) In plane
$J-\Delta$, for fixed $J_{0}=1$.}
\end{figure}

In fig.\ref{fig:Phase-diagram-zero}a is illustrated the phase diagram
$\Delta$ against $J_{0}$, for fixed $J=1$, where we observe all
five phases. The boundary between $FM$ and $PAF$ is given by $\Delta=-2J_{0}+\sqrt{4J_{0}^{2}+1}$,
analogously the interface between $PAF$ and $FRU$ is described by
$\Delta=1-2J_{0}$, similarly the boundary between $DAF$ and $AFM$
is limited by the curve $\Delta=2J_{0}+\sqrt{4J_{0}^{2}+1}$, whereas
the boundary between $AFM$ and $FRU$ is described by $\Delta=1+2J_{0}$.
In fig.\ref{fig:Phase-diagram-zero}b is depicted another phase diagram
$J$ versus $\Delta$ for a fixed parameter $J_{0}=1$. Illustrating
once again the previous phases displayed in fig.\ref{fig:Phase-diagram-zero}a,
whose boundary between $FM$ and $PAF(AFM)$ is described by the curve
$\Delta=-2+\sqrt{4+J^{2}}$, and similarly the interface between $PAF(AFM)$
and $FRU$ is given by $\Delta=-2+|J|$.

It is worth to mention that, the energy degeneracy in the boundary
of $DAF$ and $FM$ per unit cell, each dimers (Heisenberg spins)
contributes with $2^{2}$ configurations ($\vupentg,\vdpentg$) and
4 Ising spins with $2^{4}$ configurations, thus the residual entropy
is $\mathcal{S}=k_{B}\ln(2^{2N}\times2^{4N})/N=6k_{B}\ln(2)$. There
is a point for $J_{0}=0$, $J=1$ and $\Delta=1$ where all phases
coexist which is a highly frustrated phase, each dimers (Heisenberg
spins) contributes with the triplet state ($\vupentg,\vdpentg,\vtpentg$)
and 4 Ising spins ($2^{4}$) whose residual entropy is $\mathcal{S}=k_{B}\ln(3^{2N}\times2^{4N})/N=2k_{B}\ln(12)$.

Using a similar reasoning in fig.\ref{fig:Phase-diagram-zero}a, the
curve surrounding the frustrated region, becomes a frustrated curve
with a residual entropy $\mathcal{S}=k_{B}\ln(3)$, whereas the curve
limiting the boundary between $FM(DAF)$ and $PAF(AFM)$ has a residual
entropy $\mathcal{S}=k_{B}\ln(2)$.

At first glance, in fig.\ref{fig:Phase-diagram-zero}a we can observe
that for $\Delta>0$ we could have a ferromagnetic coupling, so we
should not expect a frustrated state in this region because the spins
are aligned parallel to the $z$-axis. However, we observe a frustrated
region for $\Delta>0$, because the Heisenberg spins have projections
on the $xy$ components which contributes with $-|J|/2$ in eq.\eqref{eq:E-fru},
thus generating a geometric frustration of quantum origin. Certainly,
a quantum geometric frustration effect vanishes according $J\rightarrow0$
becoming a classical geometric frustration.

Now let us compare the phase diagram of fig.\ref{fig:Phase-diagram-zero}a
with that two-dimensional Cairo pentagonal Ising model\citep{Mrojas}
which exhibits a ferrimagnetic state, for details see fig.4 of reference
\citep{Mrojas}. In the stripe of Cairo pentagonal lattice, there
is no ferrimagnetic phase, in principle the $DAF$ phase should be
the responsible state to generate the ferrimagnetic phase in the two-dimensional
lattice. It is easy to recognize the top and bottom particles will
be sharing with neighboring unit cells, then, we will have a non-zero
magnetization per unit cell, so generating a ferrimagnetic phase in
two-dimensional lattice model. While the arise of $AFM$ phase will
not be allowed in two-dimensional lattice, because the sharing particle
spin will not be equivalent between unit cells, also a similar property
forbids $PAF$ in the two-dimensional lattice model. In a nutshell,
the $AFM$ and $PAF$ phases only emerge in a one-dimensional pentagonal
chain.

We can view the Cairo pentagonal chain as a decorated orthogonal dimer
chain\citep{henrique,taras}, where in the Cairo pentagonal chain
the Ising spin would be considered as a decorated spin. Therefore,
we can compare the fig.\ref{fig:Phase-diagram-zero} with fig.3 of
reference \citep{henrique}, and we observe that both figures are
somewhat similar, particularly the phase boundaries. Although there
is a difference between them, the orthogonal dimer chain does not
exhibit a frustrated phase region\citep{henrique}, unless in the
phase boundaries.

\section{Thermodynamics of the model}

The partition function of a Cairo pentagonal Ising-Heisenberg stripe
can be obtained through the transfer matrix technique\citep{baxter}.

The Boltzmann factor for an $ab$-dimer elementary cell is given by
\begin{equation}
\mathsf{w}(s_{1,i},s_{2,i},s_{3,i},s_{4.i})={\rm tr}_{ab}\left(\text{e}^{-\beta H_{i,i}^{ab}}\right),
\end{equation}
where $\beta=1/k_{B}T$, with $k_{B}$ being the Boltzmann's constant
and $T$ is the absolute temperature.

Whereas for a $cd$-dimer the Boltzmann factor is expressed by 
\begin{equation}
\bar{\mathsf{w}}(s_{1,i+1},s_{4,i},s_{3,i},s_{2.i+1})={\rm tr}_{cd}\left(\text{e}^{-\beta H_{i,i+1}^{cd}}\right).
\end{equation}
The best way to perform the trace is to diagonalize the Hamiltonian
for $ab$-dimer and $cd$-dimer.

Using the standard 8-vertex model notation\citep{baxter} as successfully
used in two-dimensional spin lattice model\citep{urumov,Mrojas},
we can express the Boltzmann factors for $ab$-dimer as follows 
\begin{alignat}{1}
\omega_{1}= & \mathsf{w}(+,+,+,+)=z\left(x^{4}+x^{-4}\right)+\frac{y^{2}+y^{-2}}{z},\label{eq:w1}\\
\omega_{2}= & \mathsf{w}(+,-,+,-)=2z+\frac{y^{2}+y^{-2}}{z},\\
\omega_{3}= & \mathsf{w}(+,+,-,-)=\omega_{2},\\
\omega_{4}= & \mathsf{w}(+,-,-,+)=2z+\frac{y_{2}^{2}+y_{2}^{-2}}{z},\\
\omega_{5}= & \mathsf{w}(+,+,+,-)=z\left(x^{2}+x^{-2}\right)+\frac{y_{1}^{2}+y_{1}^{-2}}{z},\\
\omega_{6}= & \mathsf{w}(+,+,-,+)=\omega_{5},\\
\omega_{7}= & \mathsf{w}(+,-,+,+)=\omega_{5},\\
\omega_{8}= & \mathsf{w}(-,+,+,+)=\omega_{5},\label{eq:w8}
\end{alignat}
where $x=\mathrm{e}^{\beta J_{0}/4}$, $y=\mathrm{e}^{\beta J/4}$
and $z=\mathrm{e}^{\beta\Delta/4}$, we also define the following
exponents $y_{1}=\mathrm{e}^{\beta\sqrt{J^{2}+J_{0}^{2}}/4}$ and
$y_{2}=\mathrm{e}^{\beta\sqrt{J^{2}+4J_{0}^{2}}/4}$ just to simplify
our notation.

Analogously, the Boltzmann factors for the $cd$-dimer are expressed
in a similar way to the $ab$-dimer. Therefore, we have 
\begin{alignat}{1}
\bar{\omega}_{1}=\bar{\mathsf{w}}(+,+,+,+)= & z'\left(x^{4}+x^{-4}\right)+\frac{{y'}^{2}+{y'}^{-2}}{z'},\\
\bar{\omega}_{2}=\bar{\mathsf{w}}(+,-,+,-)= & 2z'+\frac{{y'}^{2}+{y'}^{-2}}{z'},\\
\bar{\omega}_{4}=\bar{\mathsf{w}}(+,-,-,+)= & 2z'+\frac{{y'_{2}}^{2}+{y'_{2}}^{-2}}{z'},\\
\bar{\omega}_{5}=\bar{\mathsf{w}}(+,+,+,-)= & z'\left(x^{2}+x^{-2}\right)+\frac{{y'_{1}}^{2}+{y'_{1}}^{-2}}{z'},
\end{alignat}
where $y'=\mathrm{e}^{\beta J'/4}$ and $z'=\mathrm{e}^{\beta\Delta'/4}$,
we define also the following exponents ${y'_{1}}=\mathrm{e}^{\beta\sqrt{{J'}^{2}+J_{0}^{2}}/4}$
and ${y'_{2}}=\mathrm{e}^{\beta\sqrt{{J'}^{2}+4J_{0}^{2}}/4}$. Moreover,
we also have the following relations: $\bar{\omega}_{2}=\bar{\omega}_{3}$
and $\bar{\omega}_{5}=\bar{\omega}_{6}=\bar{\omega}_{7}=\bar{\omega}_{8}$.

To study the thermodynamics of the Cairo pentagonal Ising-Heisenberg
chain, we observe that the Hamiltonian of each unit cell commutes
between them. Consequently, the partition function could be written
as the product of Boltzmann factors corresponding to the unit cells,

\begin{equation}
\mathcal{Z}_{N}=\mathrm{tr}\left(\prod_{i=1}^{N}\mathrm{e}^{-\beta(H_{i,i}^{ab}+H_{i,i+1}^{cd})}\right).\label{eq:Z_N}
\end{equation}

Therefore, the partition function \eqref{eq:Z_N} can be obtained
using the transfer matrix approach, whose transfer matrix elements
become

\begin{equation}
\mathsf{T}(s_{1},s_{2},s'_{1},s'_{2})=\sum_{s_{3},s_{4}}\mathsf{w}(s_{1},s_{2},s_{3},s_{4})\bar{\mathsf{w}}(s'_{1},s_{4},s_{3},s_{2}^{\prime}).
\end{equation}
The matrices $\boldsymbol{\mathsf{w}}$ and $\bar{\boldsymbol{\mathsf{w}}}$
corresponding to $ab$-dimer and $cd$-dimer respectively, can be
expressed explicitly by

\begin{equation}
\boldsymbol{\mathsf{w}}=\left[\begin{array}{cccc}
\omega_{1} & \omega_{5} & \omega_{5} & \omega_{3}\\
\omega_{5} & \omega_{4} & \omega_{2} & \omega_{5}\\
\omega_{5} & \omega_{2} & \omega_{4} & \omega_{5}\\
\omega_{3} & \omega_{5} & \omega_{5} & \omega_{1}
\end{array}\right],\;\bar{\boldsymbol{\mathsf{w}}}=\left[\begin{array}{cccc}
\bar{\omega}_{1} & \bar{\omega}_{5} & \bar{\omega}_{5} & \bar{\omega}_{4}\\
\bar{\omega}_{5} & \bar{\omega}_{2} & \bar{\omega}_{3} & \bar{\omega}_{5}\\
\bar{\omega}_{5} & \bar{\omega}_{3} & \bar{\omega}_{2} & \bar{\omega}_{5}\\
\bar{\omega}_{4} & \bar{\omega}_{5} & \bar{\omega}_{5} & \bar{\omega}_{1}
\end{array}\right].
\end{equation}

It is worth mentioning that the matrices $\boldsymbol{\mathsf{w}}$
and $\bar{\boldsymbol{\mathsf{w}}}$ are fully symmetric matrices.

Thus, the transfer matrix $\boldsymbol{\mathsf{T}}$ becomes

\begin{equation}
\boldsymbol{\mathsf{T}}=\boldsymbol{\mathsf{w}}\bar{\boldsymbol{\mathsf{w}}}=\left[\begin{array}{cccc}
\tau_{1} & \tau_{5} & \tau_{5} & \tau_{4}\\
\tau_{7} & \tau_{2} & \tau_{3} & \tau_{7}\\
\tau_{7} & \tau_{3} & \tau_{2} & \tau_{7}\\
\tau_{4} & \tau_{7} & \tau_{7} & \tau_{1}
\end{array}\right],
\end{equation}
where $\tau_{i}$ are the elements of transfer matrix, which are denoted
following the eight-vertex model similar to that denoted in reference
\citep{urumov,Mrojas}. Then $\tau_{i}$ are expressed by 
\begin{alignat}{1}
\tau_{1}= & \omega_{1}\bar{\omega}{}_{1}+\omega_{3}\bar{\omega}{}_{4}+2\omega_{5}\bar{\omega}{}_{5},\label{eq:tau1}\\
\tau_{2}= & \omega_{2}\bar{\omega}{}_{3}+\omega_{4}\bar{\omega}{}_{2}+2\omega_{5}\bar{\omega}{}_{5},\\
\tau_{3}= & \omega_{2}\bar{\omega}{}_{2}+\omega_{4}\bar{\omega}{}_{3}+2\omega_{5}\bar{\omega}{}_{5},\\
\tau_{4}= & \omega_{1}\bar{\omega}{}_{4}+\omega_{3}\bar{\omega}{}_{1}+2\omega_{5}\bar{\omega}{}_{5},\\
\tau_{5}= & \left(\omega_{2}+\omega_{4}\right)\bar{\omega}{}_{5}+\omega_{5}\left(\bar{\omega}{}_{1}+\bar{\omega}{}_{4}\right),\\
\tau_{7}= & \left(\omega_{1}+\omega_{3}\right)\bar{\omega}{}_{5}+\omega_{5}\left(\bar{\omega}{}_{2}+\bar{\omega}{}_{3}\right),\label{eq:tau7}
\end{alignat}
and we can also verify that $\tau_{3}=\tau_{2}$, $\tau_{5}=\tau_{6}$
and $\tau_{7}=\tau_{8}$.

Note that the transfer matrix $\boldsymbol{\mathsf{T}}$ is a non-symmetric
matrix, because in general $\tau_{5}\ne\tau_{7}$, despite each $\boldsymbol{\mathsf{w}}$
and $\bar{\boldsymbol{\mathsf{w}}}$ are perfectly symmetric.

Thus, using the transfer matrix approach the partition function \eqref{eq:Z_N},
can be written as 
\begin{equation}
\mathcal{Z}_{N}=\mathrm{tr}\left[\left(\boldsymbol{\mathsf{w}}\bar{\boldsymbol{\mathsf{w}}}\right)^{N}\right]=\mathrm{tr}\left[\boldsymbol{\mathsf{T}}^{N}\right].
\end{equation}
The eigenvalues of transfer matrix, can be obtained from $\text{det}(\boldsymbol{\mathsf{T}}-\lambda)=0$,
which results in, 
\begin{equation}
\left(\lambda^{2}-a_{1}\lambda+a_{0}\right)(\lambda-\tau_{2}+\tau_{3})\left(\lambda-\tau_{1}+\tau_{4}\right)=0,\label{eq:cubic-eq}
\end{equation}
where the coefficients of a quadratic equation are given by 
\begin{alignat}{1}
a_{1}= & \left(\tau_{1}+\tau_{2}+\tau_{3}+\tau_{4}\right),\\
a_{0}= & \left(\tau_{1}+\tau_{4}\right)\left(\tau_{2}+\tau_{3}\right)-4\tau_{5}\tau_{7}.
\end{alignat}

Consequently, the eigenvalues of the transfer matrix can be expressed
as follows 
\begin{align}
\lambda_{0}= & \tau_{2}-\tau_{3}=0,\\
\lambda_{1}= & \tau_{1}-\tau_{4},\\
\lambda_{\pm}= & \tfrac{\tau_{1}+\tau_{2}+\tau_{3}+\tau_{4}}{2}\pm\tfrac{\sqrt{\left(\tau_{1}-\tau_{2}-\tau_{3}+\tau_{4}\right)^{2}+16\tau_{5}\tau_{7}}}{2}.
\end{align}

Although, the transfer matrix $\boldsymbol{\mathsf{T}}$ is a non-symmetric
one, we can observe that all eigenvalues are obviously real functions.
Besides, one can readily identify, there is a largest eigenvalue positively
defined $\lambda_{+}$, because all $\tau_{i}$ are positive real
numbers following the relations (\ref{eq:tau1}-\ref{eq:tau7}).

In thermodynamic limit, the free energy per unit cell depends of the
the largest eigenvalue of the transfer matrix, which is expressed
by

\begin{equation}
f=-\frac{1}{\beta}\ln\left(\lambda_{+}\right).
\end{equation}

Using the free energy, we are able to obtain several thermodynamics
quantities.

\section{Physical quantities}

In what follows we will discuss the entropy ($\mathcal{S}=-\frac{\partial f}{\partial T}$)
property of the Cairo pentagonal chain, illustrating the regions where
the model exhibits a frustrated sector as well as the different antiferromagnetic
phases found in the previous section.

\begin{figure}[h]
\centering{}\includegraphics[scale=0.28]{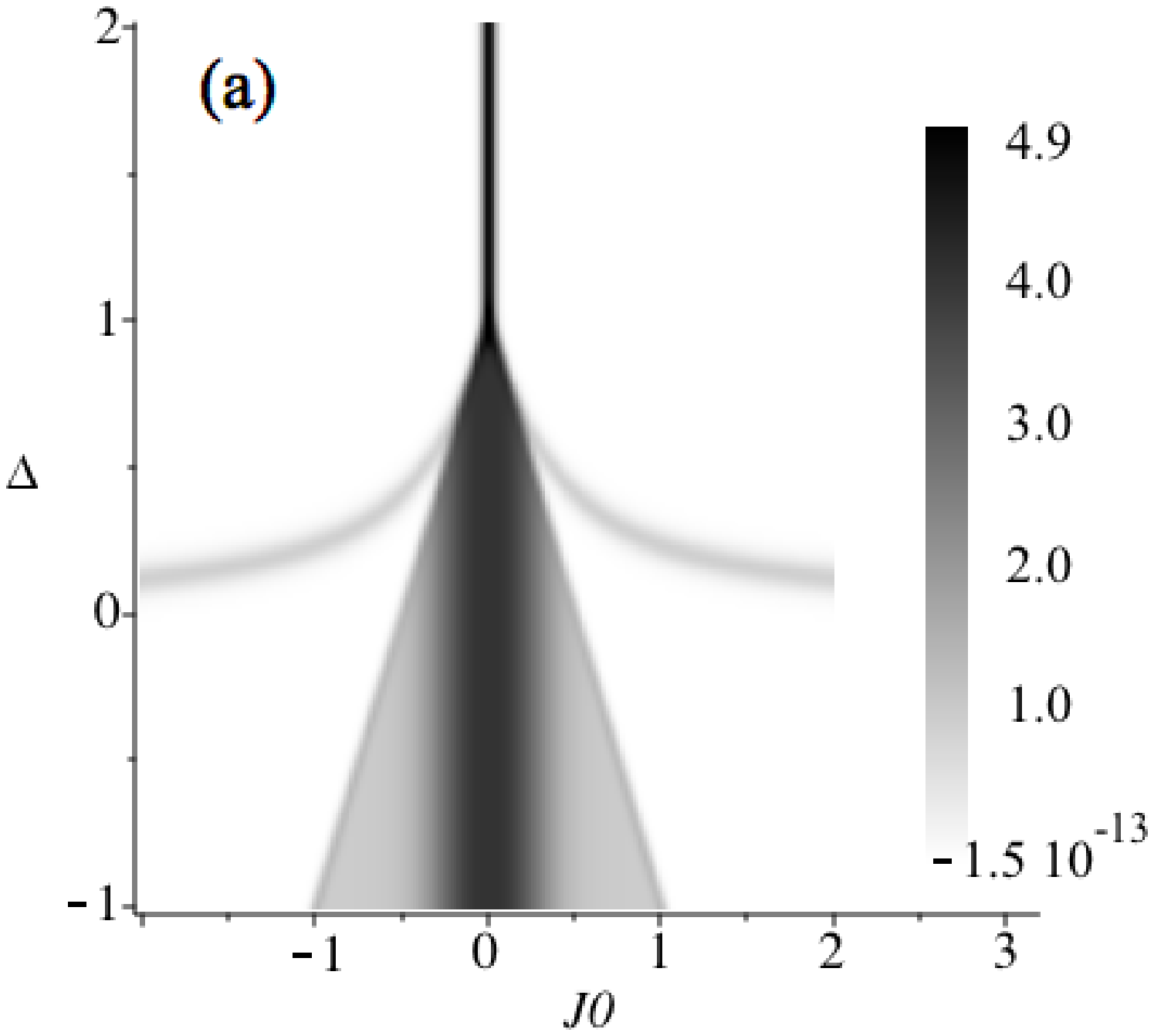}\includegraphics[scale=0.28]{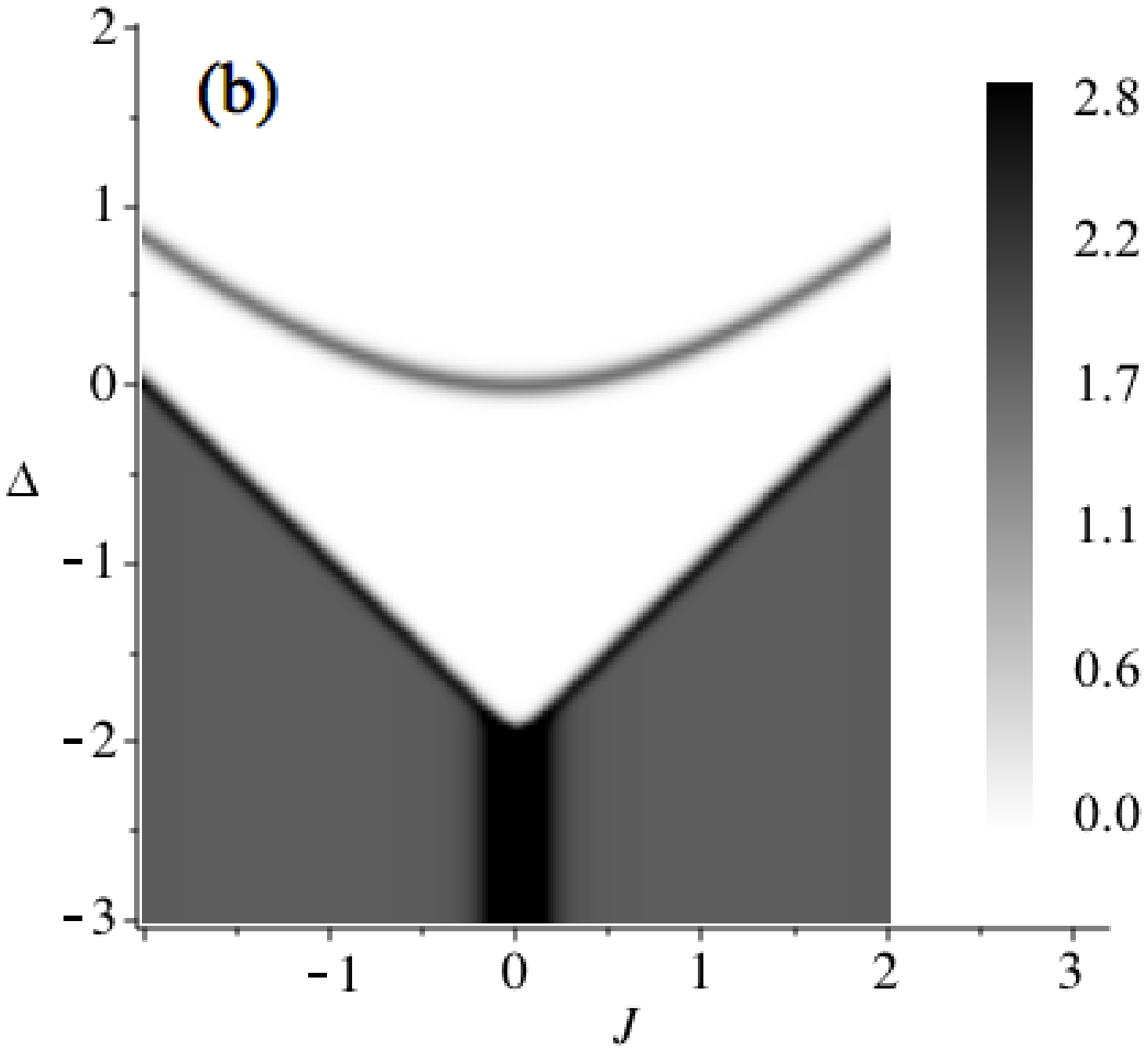}\caption{\label{fig:S-densty}(a) Density plot of entropy as a function of
$J_{0}$ and $\Delta$, assuming fixed $J=1$ and $T=0.01$. (b) Density
plot of entropy as a function of $J$ against $\Delta$ for fixed
$J_{0}=1.0$ and $T=0.01$. }
\end{figure}

In fig.\ref{fig:S-densty}a, we illustrate the density plot of the
entropy as a function of $J_{0}$ and $\Delta$ for a fixed parameter
$J=1$ and in the low-temperature limit $T=0.01$, darker regions
correspond to higher entropies. Thus, we can readily verify the evidence
of a frustrated region (FRU), with residual entropy $\mathcal{S}\rightarrow k_{B}\ln(2)$,
and the boundary of this region is also a frustrated state with residual
entropy $\mathcal{S}\rightarrow k_{B}\ln(3)$. Furthermore, there
is a frustration curve in the interface of FM and DAF (PAF and AFM)
both with residual entropy provided by $\mathcal{S}\rightarrow k_{B}\ln(2)$.
The darkest region corresponds to $J_{0}=0$, $\Delta=1$ and $J=1$,
which corresponds to a trivial frustrated phase composed by uncoupled
$ab$-dimers and $cd$-dimers, so there are 4 Ising spins per unit
cell and 2 triplet state (dimers) per unit cell, as discussed previously
then the residual entropy leads to $\mathcal{S}\rightarrow2k_{B}\ln(12)\approx4.9698$.

Similarly, in fig.\ref{fig:S-densty}b we illustrate the density plot
of entropy in the plane $J$ against $\Delta$, in the low-temperature
limit $T=0.01$ and for fixed parameter $J_{0}=1$. Our results exhibit
once again the presence of a frustrated region of the model (see fig.\ref{fig:Phase-diagram-zero}).
The darkest region corresponds to $J=0$, thus, the Cairo pentagonal
chain reduces to a pure Cairo pentagonal Ising chain with residual
entropy $\mathcal{S}=4k_{B}\ln(2)\approx2.7726$, in this region we
have an equivalent frustration to that found in reference \citep{Mrojas}.

\begin{figure}[H]
\begin{centering}
\includegraphics[scale=0.43]{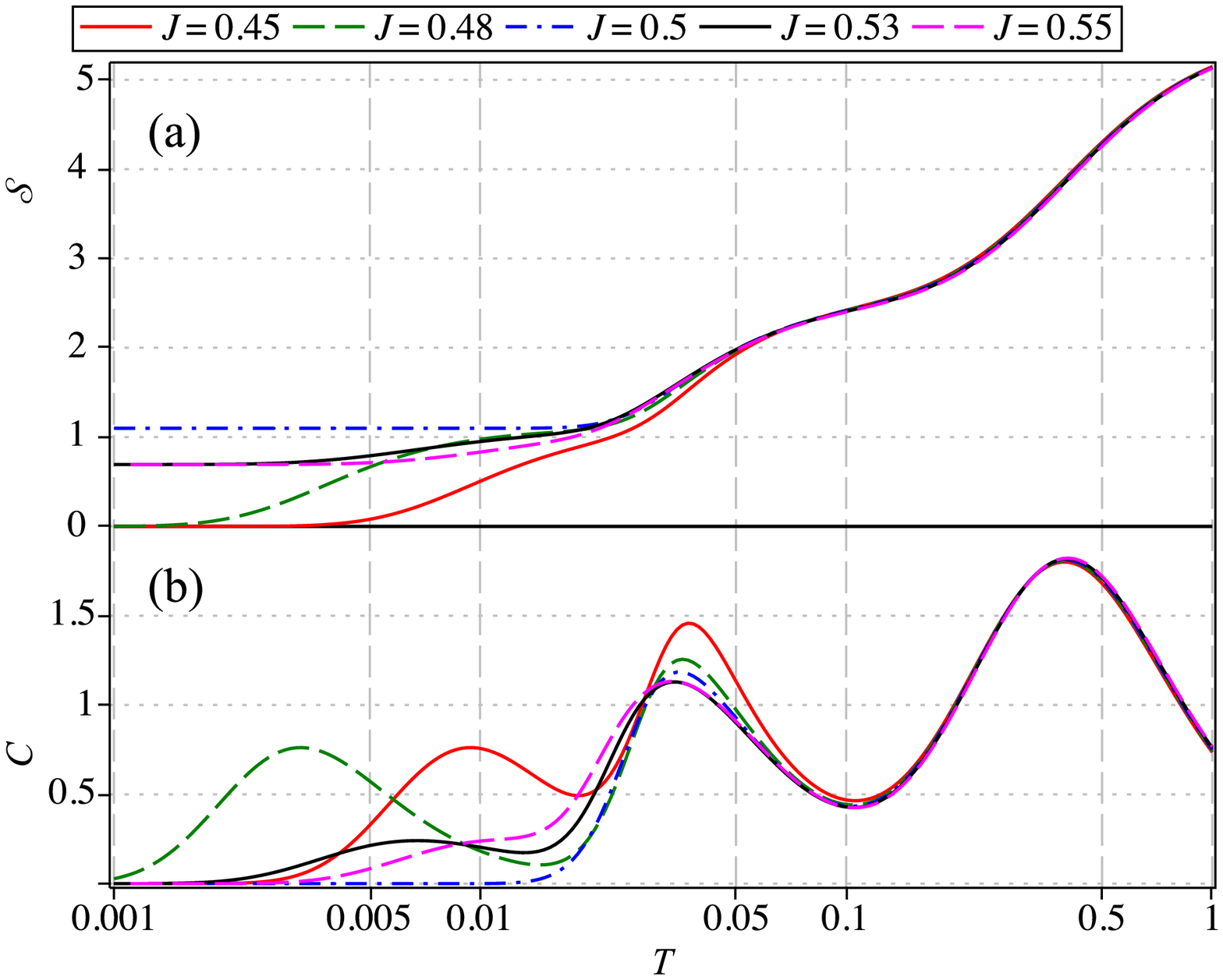}
\par\end{centering}
\centering{}\includegraphics[scale=0.43]{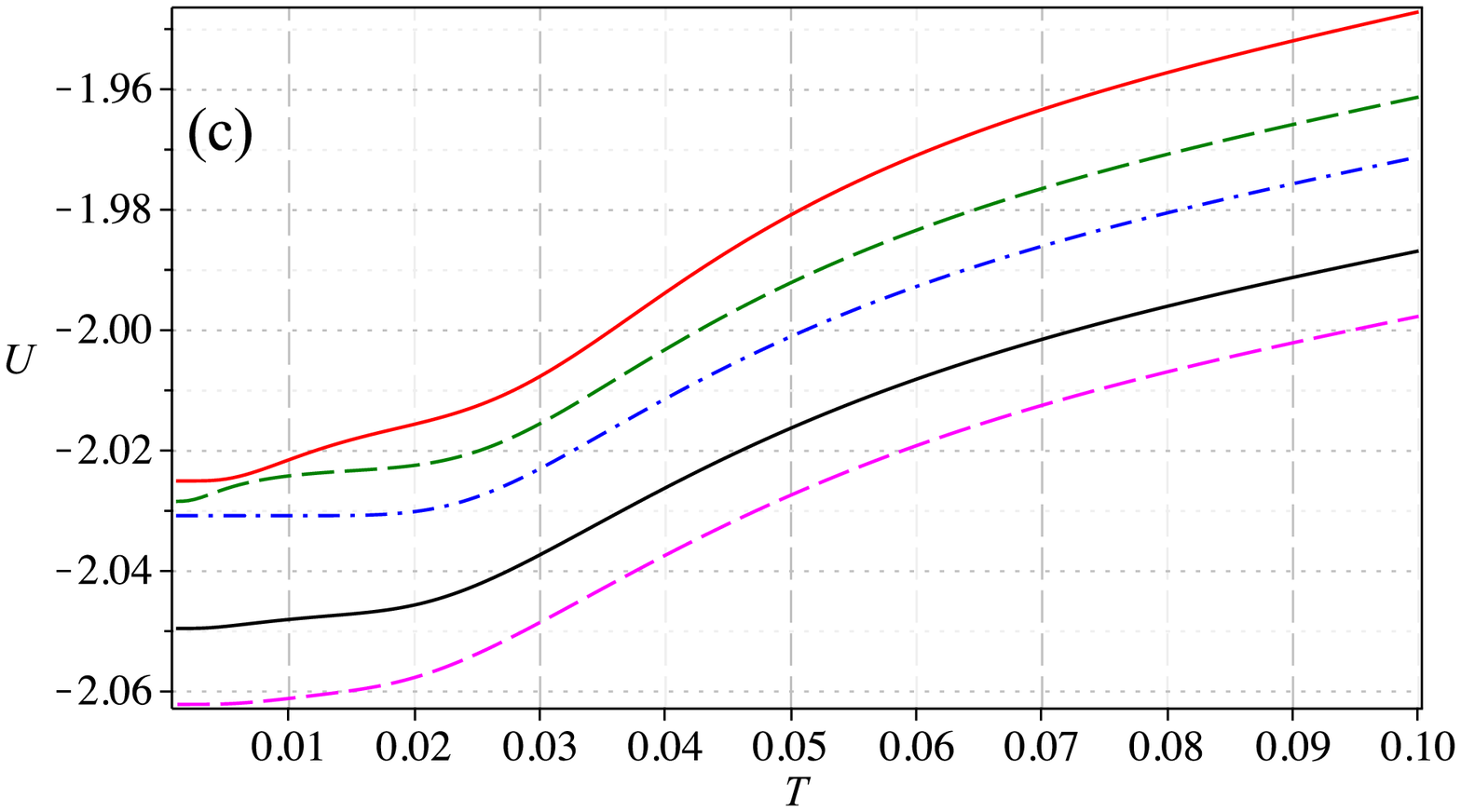}\caption{\label{fig:SCE}(a) Entropy as a function of the temperature, for
fixed values of $J_{0}=1$ and $\Delta=-1.5$, for a range of values
of $J=\{0.45,0.48,0.5,0.53,0.55\}$ in a logarithmic scale in temperature.
(b) Specific heat as a function of temperature for a same set of parameters
to the case (a). (c) Internal energy as a function of temperature
for a same set of parameters to the case of (a), but on a linear scale
for the temperature.}
\end{figure}

In fig. \ref{fig:SCE}a, we display the entropy as a function of temperature
assuming fixed parameter $J_{0}=1$ and $\Delta=-1.5$, for a range
of $J=\{0.45,0.48,0.5,0.53,0.55\}$, using conveniently a logarithmic
scale to show the low-temperature behavior, close to the phase transition
illustrated in fig. \ref{fig:Phase-diagram-zero}b. Where we show
the influence of zero temperature phase transition in the low-temperature
limit, for $|J|\apprle0.5$ the system is highly influence by PAF(AFM)
ground-state energy with residual entropy $\mathcal{S}\rightarrow0$
when $T\rightarrow0$, whereas frustrated energy contributes as low-lying
excited energy, we can observe clearly residual entropy at $J=0.5$
is given by $\mathcal{S}\rightarrow k_{B}\ln(3)$ when $T\rightarrow0$.
However for $J\apprge0.5$, the system is dominated by a frustrated
phase with residual entropy $\mathcal{S}\rightarrow k_{B}\ln(2)$
when $T\rightarrow0$.

In fig.\ref{fig:SCE}b, we also discuss another interesting thermodynamic
quantity called specific heat ($C=-T\partial^{2}f/\partial T^{2}$).
We illustrate for the same set of parameters those considered in fig.\ref{fig:SCE}a
and also on a logarithmic scale. Here we observe an anomalous double
peak in the low-temperature region, which was influenced by the zero
temperature phase transition between $PAF\leftrightarrow FRU$ and
the low-lying energy responsible for FM state\citep{katarina}. For
$J=0.45$ denoted by a solid (red) line, whose ground-state energy
is non-degenerate and the first excited energy is macroscopically
degenerate with a residual entropy $\mathcal{S}\rightarrow k_{B}\ln(2)$,
when the temperature increases the contribution of the degenerate
energy becomes more relevant than the contribution of the non-degenerate
ground state, so the entropy curve is driven to behave like a frustrated
system changing its concavity at $T\approx0.003$, whereas for specific
heat it manifests as a peak. Increasing the temperature slightly more
we observe another change of concavity in entropy at $T\approx0.03$,
this is because there is another phase transition in the neighborhood
(PAF between FM), the contribution of the excited low-lying energy
level again drives the entropy (leading to a second peak in specific
heat). A similar behavior was observed for a dashed (green) line assuming
fixed $J=0.48$ with a change of concavity in entropy at $T\approx0.01$
and a second change of concavity occurs at $T\approx0.03$, thus in
specific heat we observe a double peak. Whereas for $J=0.53$ (solid
line), the change of the concavity in entropy at the lower temperature
$T\approx0.008$ is almost imperceptible, that is because the ground
sate energy is macroscopically degenerate ($\mathcal{S}\rightarrow k_{B}\ln(2)$)
and the lowest excited energy also becomes macroscopically degenerate
($\mathcal{S}\rightarrow k_{B}\ln(3)$), this is manifest as a small
peak in the specific heat for $T\approx0.008$. While the second peak
in specific heat occurs at $T\approx0.03$ basically by the same mechanism
of low-lying excited energies. An analogous behavior we observe for
$J=0.55$ (magenta dashed line). Finally, for $J=0.5$ (doted-dashed
line) where the ground-state energy is macroscopically degenerate
($\mathcal{S}\rightarrow k_{B}\ln(3)$), and the low-lying excited
energy (that originates from FM ground-state energy) generating just
one change of concavity at around $T\approx0.03$.

In fig. \ref{fig:SCE}c, we plot the internal energy $U=f+T\mathcal{S}$,
for the same set of parameters considered in fig. \ref{fig:SCE}a,
but here, we use a linear scale just to show the low-temperature internal
energy behavior, to relate with specific heat anomalous peaks, since
$C=\partial U/\partial T$ relates both quantities.

\section{Conclusion}

In this work we have proposed the Cairo pentagonal chain, motivated
by recent discoveries of some compound such as the $\mathrm{Fe}^{3+}$
lattice in the $\mathrm{Bi}_{2}\mathrm{Fe}_{4}\mathrm{O}_{9}$ and
iron-based oxyfluoride ${\rm Bi}_{4}{\rm Fe}_{5}{\rm O}_{13}{\rm F}$
compounds with a Cairo pentagonal tiling. Therefore, we proposed one
stripe of the Cairo pentagonal Ising-Heisenberg lattice. Subsequently,
we have discussed the phase transition at zero temperature, illustrating
five phases: one ferromagnetic (FM) phase, one dimer antiferromagnetic
(DAF), one plaquette antiferromagnetic (PAF), one typical antiferromagnetic
(AFM) phase and one peculiarly frustrated (FRU) phase, where coexist
two type of frustrated states with same energy but without mixing
these phases, this kind of frustration is very unusual. It is worth
to mention also, for the case of two-dimensional pentagonal lattice
the DAF phase, will be transformed into a ferrimagnetic phase, due
to the sharing spins between the unit cells. However, the AFM and
PAF phase will be forbidden in a two-dimensional lattice, because
the sharing spins between top and bottom unit cells will not be compatible. 

To study the thermodynamics of this model we have used the transfer
matrix approach and following the eight vertex model notation to find
the partition function. Using this result, we have discussed the entropy
and specific heat as a dependence of temperature. Accordingly, we
observe an unusual behavior in the low-temperature limit, such as
residual entropy and the anomalous double peak due the existence of
three phases transition occurring in a very close region to each other
and one of them is frustrated state. Thus, the thermal excitation
of low-lying energy causes this anomalous double peak, and we also
discussed the internal energy in the low-temperature limit, where
occurred this double peak.

\section*{Acknowledgment}

F. C. R. thanks Brazilian agency CAPES for full financial support.
S. M. S. and O. R. thank Brazilian agencies, CNPq, FAPEMIG and CAPES
for partial financial support. O. R. also thanks ICTP for partial
financial support and the hospitality at ICTP.


\begin{thebibliography}{10}
\bibitem{urumov}V. Urumov, J. Phys. A: Math. Gen. \textbf{35}, 7317
(2002).

\bibitem{Mrojas}M. Rojas, O. Rojas and S. M. de Souza, Phys. Rev.
E \textbf{86}, 051116 (2012).

\bibitem{ressou}E. Ressouche, V. Simonet, B. Canals, M. Gospodinov,
V. Skumryev, Phy. Rev. Lett. \textbf{103}, 267204 (2009).

\bibitem{ralko}A. Ralko, Phys. Rev. B \textbf{84}, 184434 (2011).

\bibitem{Rouso}I. Rousochatzakis, A. M. Läuchli and R. Moessner,
Phys. Rev. B \textbf{85}, 104415 (2012).

\bibitem{pchelkina}Z.V. Pchelkina and S.V. Streltsov, Phys. Rev.
B \textbf{88}, 054424 (2013).

\bibitem{abakumov}A. M. Abakumov, D. Batuk, A. A. Tsirlin, C. Prescher,
L. Dubrovinsky, D. V. Sheptyakov, W. Schnelle, J. Hader- mann, and
G. V. Tendeloo, Phys. Rev. B \textbf{87}, 024423 (2013).

\bibitem{Isoda}M. Isoda, H. Nakano, and T. Sakai, J. Phys. Soc. Jpn.
\textbf{83}, 084710 (2014).

\bibitem{nakano}H. Nakano, M. Isoda and T. Sakai, J. Phys. Soc. Jpn.
\textbf{83}, 053702 (2014).

\bibitem{Rozova}M. G. Rozova, V. V. Grigoriev, I. A. Bobrikov, D.
S. Filimonov, K. V. Zakharov, O. S. Volkova, A. N. Vasiliev, E. V.
Antipov, A. A. Tsirlin, and A. M. Abakumov, Dalton Trans. \textbf{45},
1192 (2016).

\bibitem{Ma}Y. Ma, L. Kou, X. Li, Y. Dai, and T. Heine, NPG Asia
Mater \textbf{8}, e264 (2016).

\bibitem{chainani}A. Chainani, K. Sheshadri, arXiv:1412.6944

\bibitem{Zhang}S. Zhang, J. Zhou, Q. Wang, X. Chen, Y. Kawazoe, and
P. Jena, Proc. Natl. Acad. Sci. USA \textbf{112}, 2372 (2015).

\bibitem{shastry}B. S. Shastry and B. Sutherland, Physica B \textbf{108},
1069 (1981).

\bibitem{ivanov}N. B. Ivanov and J. Richter, Phys. Lett. A \textbf{232},
308 (1997).

\bibitem{henrique}H. G. Paulinelli, S. M. de Souza, O. Rojas, J.
Phys.: Condens. Matter \textbf{25}, 306003 (2013)

\bibitem{taras}T. Verkholyak, J. Strecka, Phys. Rev. B \textbf{88},
134419 (2013); Acta Phys. Pol. A \textbf{126}, 22 (2014); arXiv:1607.08457

\bibitem{braz}F.F. Braz, F.C. Rodrigues, S.M. de Souza, O. Rojas,
Annals of Physics \textbf{372}, 523 (2016).

\bibitem{baxter}R.J. Baxter, Exactly Solved Models in Statistical
Mechanics, (Academic Press, New York, 1982).

\bibitem{DDT}O. Rojas and S. M. Souza J. Phys. A: Math. Theor. \textbf{44},
245001 (2011).

\bibitem{dec-trans}M. E. Fisher, Phys. Rev. \textbf{113}, 969 (1959);
J. Strecka, Phys. Lett. A \textbf{374}, 3718 (2010); O. Rojas, J.
S. Valverde and S. M. de Souza, Physica A \textbf{388}, 1419 (2009).

\bibitem{katarina}K. Karlova, J. Strecka, T. Madaras, Physica B \textbf{488},
49 (2016).
\end{thebibliography}
\end{document}